\documentclass[a4paper,11pt]{article}
\pdfoutput=1 
\RequirePackage{snapshot}  

\usepackage{jheppub} 

\usepackage{bm,amsmath,amssymb,slashed,graphicx,%
            enumerate,alltt,xspace,multirow,xcolor,mathrsfs}
\usepackage{fancyvrb}
\usepackage{booktabs,tabularx}
\usepackage{graphicx}
\usepackage{subcaption}
\usepackage{xspace}
\usepackage{cancel}
\usepackage[utf8]{inputenc} 
\usepackage[compat=1.0.0]{tikz-feynman}
\usepackage{scalerel}
\usepackage{layouts}
\usepackage{adjustbox}
\usepackage{pdflscape}
\usepackage[normalem]{ulem} 
\usepackage{printlen}
\usepackage{wasysym}

\usepackage{listings}
\usepackage{algorithm}
\usepackage[noEnd=true]{algpseudocodex}

\newcommand{\UGRaff}{Departamento de F\'{i}sica Te\'{o}rica y del Cosmos, Universidad de Granada,
Campus de Fuentenueva, E-18071 Granada, Spain}
\newcommand{\CNRSaff}{Université Paris-Saclay, CNRS, CEA, Institut de physique théorique, 91191, Gif-sur-Yvette, France}
\newcommand{\MITaff}{Massachusetts Institute of Technology, Laboratory for Nuclear Science, Cambridge, MA, 02139}

\newcommand{\zmin}{z_{\rm{min}}}
\newcommand{\zmax}{z_{\rm{max}}}
\newcommand{\ptmin}{p_{t,\rm{min}}}
\newcommand{\ptmax}{p_{t,\rm{max}}}
\newcommand{\py}{{\textsc{pythia}}8\xspace}
\newcommand{\hw}{{\textsc{herwig}}7\xspace}

\title{Secondary Lund jet plane as a gluon-enriched sample}

\author[a]{Cristian Baldenegro,}
\author[b]{Alba Soto-Ontoso,}
\author[c]{Gregory Soyez}

\affiliation[a]{\MITaff}
\affiliation[b]{\UGRaff}
\affiliation[c]{\CNRSaff}

\date{Received: date / Accepted: \today}

\abstract{We propose a new strategy to obtain a high-purity sample of gluon-initiated jets at the LHC. Our approach, inspired by the Lund jet plane picture, is to perform a dijet selection where the two jets are collinear to each other and their momentum fraction share is highly asymmetric, and to measure the primary Lund plane density of emissions of the subleading jet. The subleading jet in this topology is practically equivalent to a secondary Lund jet plane. We demonstrate by means of fixed-order calculations that such a simple setup yields gluon jet fractions of around 90\% for the subleading jet for both quark- and gluon-initiated jets. This observation is confirmed using hadron-level Monte Carlo generated events. We also show that the extracted gluon purities are highly resilient to the overall colour structure of the event, to the flavour of the hard-scattering process, and to the parton distribution functions. This strategy is well-suited for constraining the radiation pattern of gluon-initiated jets using a set of fiducial cuts that can readily be tested at the LHC, without relying on taggers or statistical demixing.}

\begin{document}
\maketitle

\section{Introduction}

Jets, the collimated showers of particles produced in high-energy particle collisions, are multi-scale probes of the strong interaction. The description of the formation of jets involves a combined understanding of the hard scattering producing energetic quarks and gluons, the parton showering driving the jet evolution, as well as the transition from the collection of partons to the hadrons that are eventually detected experimentally. In high-energy proton-proton collisions, the underlying event activity, multi-parton interactions, and initial-state radiation also contribute to the transverse momentum, $p_t$, and substructure of jets.

Although jet physics can be understood analytically to a great extent, in practice Monte Carlo (MC) event simulation is often used for phenomenological and experimental applications. For jet physics specifically, MC event simulation provides the baseline in searches for new physics, is a key ingredient in the calibration and flavour tagging of such jets, and is used for precision tests of Quantum Chromodynamics (QCD)~\cite{Campbell:2022qmc}.

A longstanding problem in MC event generators is the correct description of the fragmentation of gluon-initiated jets. As a paradigmatic example, the leading systematic uncertainty in the jet energy scale at the LHC is the MC uncertainty on the gluon-jet response~\cite{ATLAS:2020cli,CMS:2016lmd, CMS-DP-2020-019}.\footnote{Recently, the ATLAS collaboration has published a series of paper in which they investigate the origin of the gluon-jet response uncertainty in MCs~\cite{ATL-PHYS-PUB-2022-021,ATLAS:2024kkj, ATLAS:2024png}. Their results indicate that the baryon fraction inside jets play a significant role in their detector response.}  Along these lines, Refs.~\cite{Andersen:2016qtm, Gras:2017jty} showed that there is a much wider spread in predictions for gluon showers than for quark fragmentation for a number of jet substructure observables. This is mostly because MC generators are tuned to describe hadronic final states produced in $e^+e^-$ collisions as a baseline, so that quark radiation patterns are much more strongly constrained than gluon ones. The mismodeling of gluon jet showers trickles down to uncertainties in other quantities at the LHC as well as the size of the calibration factors for machine-learning-based quark-gluon discriminators, which rely on MC generated events for training~\cite{ATLAS:2014vax}. Having access to a gluon-enriched sample would reduce these uncertainties.\footnote{Throughout this paper we use a definition of gluon-enrichment in the spirit of Ref.~\cite{Andersen:2016qtm, Gras:2017jty}, namely ``phase-space region (as defined by an unambiguous hadronic fiducial cross section measurement) that yields an enriched sample of gluons (as interpreted by some suitable, though fundamentally ambiguous, criterion)".}

A high-purity gluon-initiated jet sample would also facilitate the extraction of the strong coupling $\alpha_s$ from jet substructure observables at the LHC. This is so since at lowest order in perturbation theory jet substructure observables are sensitive to $\alpha_s C_i$, where $C_i$ is the colour charge of the emitter. In other words, there is a degeneracy between $\alpha_s$ and the quark versus gluon fraction of jets~\cite{Proceedings:2018jsb}.\footnote{There are different ways of eliminating this degeneracy. On the one hand, there are certain observables that are free of this degeneracy, e.g. ratios of energy-energy correlators~\cite{Chen:2020vvp,CMS:2024mlf,Lee:2024esz}. Another way is to calculate the quark and gluon fractions using perturbative QCD as was done in Ref.~\cite{Hannesdottir:2022rsl} for the SoftDrop jet mass, or to include subleading corrections which partially break the degeneracy.} More generally, a strategy that allows for constraining final-state radiation of gluon-initiated jets in an isolated way would be beneficial for jet substructure studies and measurements at the LHC.

Crafting a gluon-dominated jet sample has been an active topic of research. In $e^+e^-$ collisions, a sample enriched in gluon-initiated jets can be obtained in events with two $b$-jets recoiling from a third jet, which is used as a proxy for gluon jets. Such an event topology has been measured at LEP to constrain soft gluon jets~\cite{OPAL:1991ssr,OPAL:1993uun,ALEPH:1995oxo, OPAL:1995ab, DELPHI:1995nzf, TOPAZ:1997kkz, OPAL:1999jkz, DELPHI:1999alp, CLEO:1992fdq, Derrick:1985du, ALEPH:1994hlg, OPAL:2001ekk, OPAL:2004prv, OPAL:2003wca} and has been used in recent tuning campaigns of \hw~\cite{Reichelt:2017hts,Mo:2017gzp}. In such measurements, anti-$b$ tagging on the third jet and additional topological requirements were shown to improve the gluon purities. The statistical uncertainty of this channel would be highly reduced at the future FCC-$ee$, which will also enable constraining gluon-initiated radiation by using associated Higgs boson production~\cite{Soyez:2017cwe,dEnterria:2021xij}.

At the LHC, a general strategy proposed in Refs.~\cite{Gallicchio:2011xc,Andersen:2016qtm} consists of targeting process-based and phase-space-based selections to enrich samples in quark-like or gluon-like radiation patterns. An example of the first category would be to use inclusive dijet (gluon dominated) and boson-tagged jet (quark dominated) samples to constrain quark and gluon jet substructure simultaneously. However, the interpretation of this data in the context of MC generator tuning is convolved with other effects unrelated to final-state radiation (e.g., the choice of parton distribution function (PDF), the order of the perturbative QCD (pQCD) calculation used for the hard scattering, and others), such that it becomes difficult to isolate the effect that comes from the mismodeling of gluon-initiated jets or other mechanisms in the collision. Regarding phase-space-based selections, one could rely on an inclusive jet sample and change the relative quark/gluon fraction by exploring different jet rapidities~\cite{ATLAS:2019mgf, Aad:2019onw, Pablos:2022mrx} or centre-of-mass energies~\cite{Baron:2023hkp}. 

A conceptually different approach to the ones discussed above pursues quark/gluon discrimination on a jet-by-jet basis exploiting the simple idea that gluons radiate more than quarks. This can be done using jet substructure observables~\cite{Krohn:2012fg,Frye:2017yrw,Kang:2021ryr,Stewart:2022ari}, machine learning strategies~\cite{Komiske:2016rsd,Dreyer:2020brq,Dolan:2022ikg}, or a combination of both~\cite{Dreyer:2021hhr}. Yet another strategy consists of using statistical demixing, for example with topic modeling, to separate quark and gluon distributions at the event ensemble level ~\cite{Larkoski:2014pca,Metodiev:2018ftz,Komiske:2018vkc,ATLAS:2019mgf,Aad:2019onw,LeBlanc:2022bwd,Komiske:2022vxg}. In the latter, the connection with the partonic degrees of freedom is not as straightforward.

In this paper, we present a strategy to create a gluon-enriched sample based on the Lund jet plane picture~\cite{Andersson:1988gp,Dreyer:2018nbf}. The Lund jet plane is a tool to create a proxy for the actual parton shower using the jet clustering tree. First, it re-organises the jet into an angular ordered sequence using the Cambridge/Aachen (C/A) algorithm~\cite{Dokshitzer:1997in}. Then, it proceeds by unwinding the C/A clustering sequence. The first declustering, i.e., the pair of subjets at largest angles, can be viewed as a proxy for the \textit{first} emission in the shower (in an angular-ordered picture). 
The \textit{primary} declusterings are constructed by iteratively undoing the C/A clustering sequence recursing, at each step, into the hardest of the two subjets, where, in a hadron collider context, hardest means having the largest transverse momentum with respect to the beam, $p_t$. The set of softer subjets along this chain is referred to as the \textit{primary emissions}.
This procedure can be repeated deeper into the C/A clustering tree. For each of the primary declusterings, we can do the same iterative declustering exercise, yielding the \textit{secondary} declusterings and emissions associated with this primary declustering. To each of the secondary declusterings we could then associate \textit{ternary} declusterings, and so on.

One can then build observables out of the kinematics of these declusterings. The simplest one is the primary Lund plane density, i.e.,\ the density of primary emissions as a function of their relative transverse momentum and opening angle. The primary Lund plane density has been recently measured in inclusive jets~\cite{ATLAS:2020bbn,CMS:2023lpp} and boosted top quark jets \cite{ATLAS:2024dua} at the LHC and compared to first-principle calculations~\cite{Lifson:2020gua} and MC simulations. It has been shown that this observable is a powerful tool to constrain hadronisation and parton showers in a factorised way. Recently, the Lund plane picture has been used to define a multiplicity-like observable that is both infrared and collinear safe and amenable to precision calculations~\cite{Medves:2022ccw,Medves:2022uii,ATLAS:2024wrd}. Regarding quark/gluon discrimination, Ref.~\cite{Dreyer:2021hhr}
computed analytically a likelihood discriminator using either the primary Lund plane or the full clustering tree and compared it to machine learning tools trained with the same input. 

Here, we propose to create a gluon-enriched jet sample using the secondary Lund plane and, more specifically, to constrain gluon-initiated radiation patterns via the secondary Lund plane density. Recalling that the secondary emissions are defined as emissions from each of the primary emissions, i.e.,\ each of the softer branches in the primary declusterings,  one expects from perturbative QCD that the softest branch will often be initiated by a gluon, due to the soft singularity of the splitting function. This will not always be the case, and part of this paper is devoted to exploring different definitions of the secondary Lund plane density aiming at enhancing the gluon fraction. One of such possibilities is to construct the secondary Lund plane density by using the subleading jet in a dijet pair after imposing some pQCD-inspired fiducial cuts. We show that the subleading jet is initiated by a gluon up to 90\% of the times.

The secondary Lund plane density has been used in other contexts. Ref.~\cite{Dreyer:2018nbf} used it to improve $W$ boson jet tagging performance at lower jet $p_t$. Secondary Lund jet planes have also been used by the ATLAS Collaboration in graph neural networks to separate boosted $W$ bosons from the QCD background~\cite{ATL-PHYS-PUB-2023-017}. Another type of secondary Lund planes has been used to improve the data-to-MC agreement for multiprong jets, as presented by the CMS Collaboration~\cite{CMS-DP-2023-046}.

This paper is organized as follows. First we present at the beginning of Sec.~\ref{sec:rhos-def} the algorithm to construct the secondary Lund plane density as introduced in Ref.~\cite{Dreyer:2018nbf}. Next, we discuss a few experimentally feasible strategies that target an enhanced gluon fraction in Sec.~\ref{sec:enhance-gfrac}. For each of the methods, we compute in Sec.~\ref{sec:analytics} the gluon fraction at tree level. We expand upon this proof-of-concept analysis in Sec.~\ref{sec:MCs}, where we perform a series of MC studies that demonstrate the robustness of the extracted gluon purities in terms of the secondary Lund jet plane density. We conclude and discuss some applications of our proposed strategy in Sec.~\ref{sec:end}.    
\section{The secondary Lund plane density}
\label{sec:rhos-def}

For a given jet (or subjet) $j$, we introduce a recursive declustering procedure following the `hardest' branch as follows:
\begin{itemize}
  \item start with $j_1=j$,
  \item at step $i$, we undo the last clustering of $j_i$ into a pair of subjets $j_i^\text{(hard)}$ and $j_i^\text{(soft)}$, such that $p_{ti}^\text{(hard)}>p_{ti}^\text{(soft)}$. We record the \textit{Lund coordinates} $\mathcal{T}_i\equiv\{\Delta_i,k_{ti},z_i,\dots\}$ as the kinematics of this declustering, i.e.,
  \begin{subequations}\label{eq:z-def}
  \begin{align}
    \Delta^2_i & = (y_i^\text{(soft)}-y_i^\text{(hard)})^2 + (\phi_i^\text{(soft)}-\phi_i^\text{(hard)})^2 , \\
    k_{ti} &= p_{ti}^\text{(soft)}\Delta_{i},   
    \qquad z_i = \frac{p_{ti}^\text{(soft)}}{p_{ti}^\text{(hard)}+p_{ti}^\text{(soft)}},
  \end{align}
  \end{subequations}
  where all variables are measured with respect to the colliding beams. 
  \item define $j_{i+1}=j_i^\text{(hard)}$ and iterate the procedure until no declusterings are left.
\end{itemize}
This procedure defines an ordered list of \textit{Lund emissions} and associated \textit{Lund coordinates} 
\begin{equation}
 \mathcal{L}^\text{(emis)}(j)\equiv [j_1^\text{(soft)}, \dots, j_n^\text{(soft)}]
 \qquad\text{and}\qquad \mathcal{L}^\text{(coords)}(j)=[\mathcal{T}_1,\dots,\mathcal{T}_n].
 \end{equation}
The average density of Lund emissions per jet, also known as the Lund plane density, is then given by
\begin{equation}
\label{eq:LP-density}
\rho \equiv \frac{1}{N_{\rm jets}}\frac{d\mathcal{L}^\text{(emis)}(j)}{d\ln k_t d \ln (1/\Delta)}.
\end{equation}

\begin{figure}
    \centering
    \includegraphics[width = 1\textwidth]{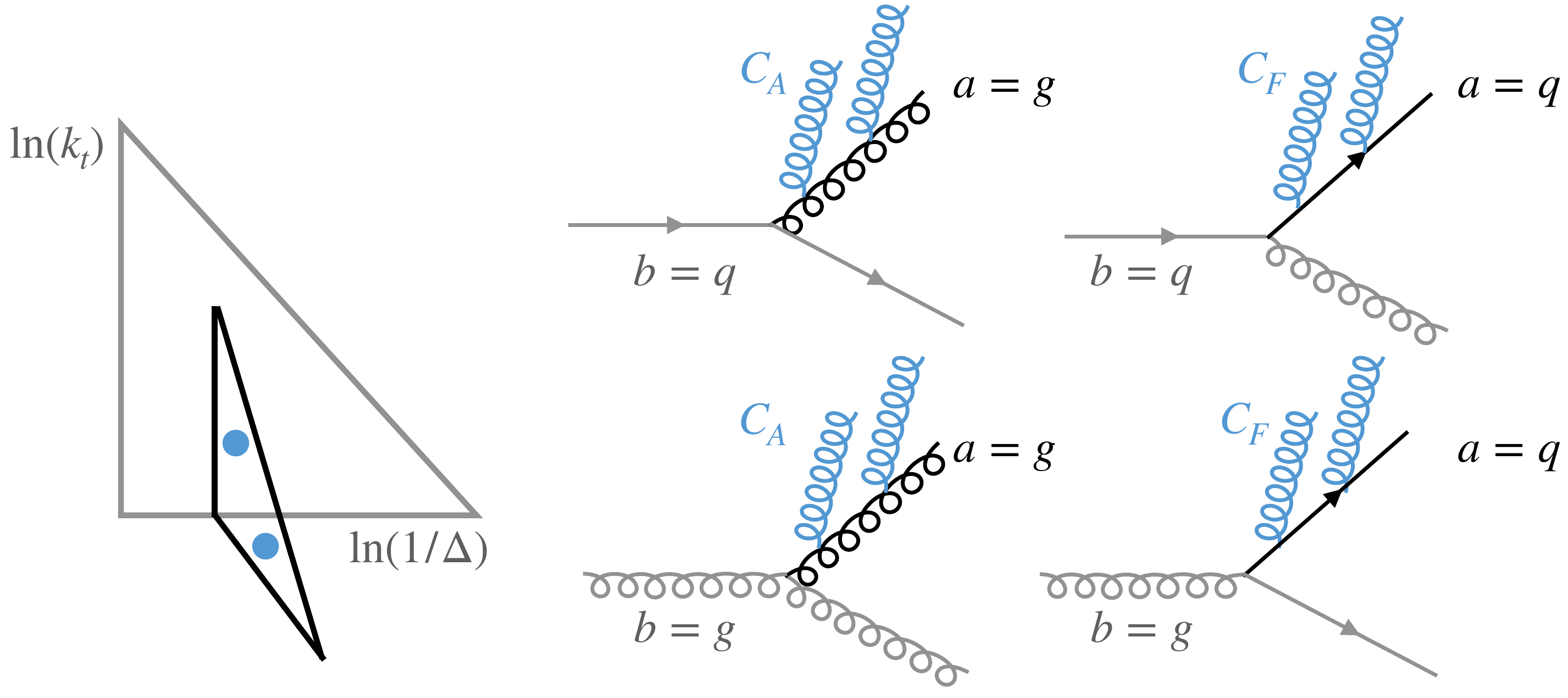}
    \caption{Schematic diagram showing the tree-level contributions to the secondary Lund jet plane density. The two leftmost diagrams correspond to emissions that spawn a secondary Lund plane enriched in gluon-initiated radiation. The two rightmost diagrams represent the two dominant contributions to parton flavour changes, where the secondary Lund plane is dominated by quark-initiated radiation.}
    \label{fig:gluonVsQuarkParisi}
\end{figure}

When applying this recursive declustering procedure to a reconstructed jet, $j$, in proton-proton collisions one first needs to recluster it with the C/A algorithm. Then, one can directly define the list of \textit{primary} emissions and Lund coordinates as 
\begin{equation}\label{eq:primaries}
\mathcal{L}_\text{primary}^\text{(emis,coords)}(j) \equiv \mathcal{L}^\text{(emis,coords)}(j).
\end{equation}
For each primary emission $j
^{\text{soft}}$, the \textit{secondary} emissions and coordinates are defined as the primary declusterings of $j
^{\text{soft}}$
\begin{equation}\label{eq:secondaries}
\mathcal{L}_{\text{secondary}}^\text{(emis,coords)}(j,j^{(\text{soft})}) \equiv \mathcal{L}_\text{primary}^\text{(emis,coords)}(j^\text{(soft)}).
\end{equation}
At this stage, one can either apply Eq.~\eqref{eq:secondaries} to all primary emissions or to just one declustering chosen following an infrarred-and-collinear (IRC) safe procedure. For the purpose of this paper, it is sufficient to focus on the latter. The key observation is that secondary emissions are mostly gluons due to the singularity structure of QCD amplitudes. To illustrate this point, let us work at first order in the strong coupling constant, $\alpha_s$, and in the soft-and-collinear limit, i.e., the splitting/declustering satisfies $\Delta \ll R, k_t \ll p_{t,{\rm jet}}$, where $R$ is the reconstructed jet cone radius. At this level of accuracy, the density of secondary emissions is given by 
\begin{equation}\label{eq:rhos-double-log}
\rho_s \equiv \frac{1}{N_{\rm jets}}\frac{d\mathcal{L}_{\text{secondary}}^\text{(emis)}(j,j^{(\text{soft})})}{d\ln k_t d \ln (1/\Delta)} \sim \frac{2\alpha_s(k_t)C_R}{\pi} 
\end{equation}
where $C_R$ is the colour factor of the first, secondary emission. The value of $C_R$ fluctuates on a jet-by-jet basis depending on both the flavour of the jet-initiator and the kinematics of the splitting, which will determine which branch is tagged as secondary. Our goal is to maximize the configurations in which $C_R=C_A$. At leading order, there are four possibilities to consider, as displayed in Fig.~\ref{fig:gluonVsQuarkParisi}. Let us first discuss the case in which the parton initiating the jet is a quark, such that the first splitting is always $q\to qg$. Since the $P_{gq}$ splitting function has a soft divergence, the gluon will take, on average, a small fraction of the parent's energy and will thus be tagged as secondary, i.e., $C_R=C_A$. This is spoiled by splittings with $z>1/2$, since then the quark is tagged as secondary and $C_R=C_F$. For gluon-initiated jets, there are two possible splitting channels: $g\to gg$ and $g\to q\bar q$. The former needs no discussion since the secondary Lund plane is always spanned by a gluon so that $C_R=C_A$. In turn, the latter will always contribute with a $C_R=C_F$ factor. It is important to note that this splitting channel is naturally suppressed since it does not have a soft singularity. Thus, at leading-order, the $C_R=C_A$ case is favored by the soft singularity of QCD. 

\subsection{Strategies to enhance the gluon fraction}
\label{sec:enhance-gfrac}

So far, we have kept the discussion at the level of one emission. Further considerations are required when considering multiple emissions and when gauging the experimental feasibility of such an observable. One of them is having sufficient phase-space so as to populate the secondary Lund plane. This imposes that the reconstructed jet must have  high-$p_t$ and that the emission that spawns the secondary Lund plane cannot be too soft or too collinear. The latter consideration is also important to mitigate distortions due to boundary effects, i.e., when two subjets or two jets are too close with each other. The large-angle regime is also problematic from the perspective of interpreting the results in the context of MC event generators, since they work best in the collinear limit. 

With these general criteria in mind, one can consider at least three setups to enhance the gluon fraction (with selections considering Run-2 or Run-3 high-pileup conditions):

\begin{itemize}
    \item \textbf{Grooming setup}: run anti-$k_t$ clustering \cite{Cacciari:2008gp} on the event with large jet radius $R = 1.2$. Select all jets with $p_t > 1 $ TeV and apply SoftDrop grooming with $z_{\rm{cut}} = 0.1$ and $\beta = 0$~\cite{Dasgupta:2013ihk,Larkoski:2014wba}. For a given jet, we denote $j_{\text{SD}}$ the primary declustering at largest angle that satisfies $z_{\text{min}} <z_g< z_{\text{max}}$ and $\Delta > \Delta_{\text{min}}$. If no declustering is found we discard the jet. We then apply Eq.~\eqref{eq:secondaries} with $j_{\text{SD}}=j^{\text{(soft)}}$ and obtain the secondary Lund plane density normalized to the number of jets that satisfy the above set of fiducial cuts.  
    As we argue below, the optimal values of the free parameters are $z_{\text {min}}=0.1$, $z_{\text {max}}=0.2$ and $\Delta_{\text{min}}=0.8$.
    \item \textbf{Trimming setup}: run anti-$k_t$ clustering on the event with large jet radius $R = 1.2$. Select all jets with $p_t > 1 $ TeV. Recluster each jet with a smaller $R$ (e.g., $R = 0.4$) using a clustering sequence of the (generalised) $k_t$ algorithm family. Select the two most energetic subjets, $j_{\text {lead}}$, $j_{\text{sublead}}$ and check whether they satisfy $0.1 < z < 0.2$ and $\Delta > 0.8$. If so, compute the ``secondary'' Lund plane density using $j=j_{\text{sublead}}$ in Eq.~\eqref{eq:LP-density}. If not, discard the jet. 
    \item \textbf{Dijet selection (nominal)}: run anti-$k_t$ clustering on the event with $R = 0.4$, as is the standard at the LHC. Select all jet pairs in the event where the leading jet has $p_{t,{\rm lead}} > 700$ GeV and the subleading jet has $p_{t,{\rm min}} < p_{t,{\rm sublead}} < p_{t,{\rm max}}$.\footnote{We have checked that a selection on the momentum imbalance, $z$, between the two jets also works.} We also impose an angular constraint between the two jets $\Delta_{\rm min} < \Delta R < \Delta_{\rm max}$. Specifically, as we will discuss below, we use  $p_{t,{\rm min}}=150$ GeV, $p_{t,{\rm max}}=200$ GeV, $\Delta_{\rm min} = 1$ and $\Delta_{\rm max} = 1.2$.
    We build the Lund plane density using $j\equiv j_\text{sublead}$ in Eq.~\eqref{eq:LP-density}.
\end{itemize}

The SoftDrop setup is conceptually closer to the construction of the primary/secondary Lund plane density~\cite{Dreyer:2018nbf}, since it relies on the same C/A clustering jet tree. The trimming~\cite{Krohn:2009th} setup mitigates distortions due to boundary effects at large angles. The third approach has the advantage that it relies on the standard anti-$k_t$ clustering sequence that is used by LHC experiments, and for which flavour taggers and jet energy calibrations are derived. Before delving into a dedicated MC study of the gluon purities achieved with these strategies, we provide next an analytic study at fixed-order.   

\subsection{First-principles estimates of gluon purities}
\label{sec:analytics}
The secondary Lund plane density is calculable in perturbative QCD. For the purpose of this paper, we limit ourselves to a simple first-order calculation. We postpone a fully resummed prediction for $\rho_s$ to a dedicated study.\footnote{The core pieces of the resummed predictions for the primary Lund plane density~\cite{Lifson:2020gua} can be reused for the secondary Lund plane density.} 
More specifically, we focus on the double-logarithmic approximation where the Lund-plane density, given by Eq.~\eqref{eq:rhos-double-log}, only depends on the colour factor of the parton emitting soft-collinear radiation in the secondary Lund plane.
In this setup, we compute the fraction of events where this leading parton is a gluon at first order in $\alpha_s$. This is well-defined at this order of perturbation theory and serves to illustrate that our procedure yields high-purity gluon samples. 
At this level of accuracy, the trimming setup is equivalent to the dijet selection. We thus present results for the grooming and dijet setups both at exact $\mathcal{O}(\alpha_s)$ using \texttt{NLOJet++}~\cite{Nagy:2003tz} and using a fully-analytic approach valid in the collinear limit. 
This second approach is meant to study the extent to which the gluon purity depends on the flavour of the initial hard parton, the one that triggered the splitting.

Let us begin by discussing the collinear limit. In the grooming setup, the fiducial cuts presented in the previous section amount to selecting momentum fractions between some $\zmin$ and some $\zmax$. We therefore define
\begin{equation}
\label{eq:int-Pab}
I^\text{(groom)}_{ab} \equiv I^\text{(groom)}_{ab}(\zmin,\zmax) = \int_{\zmin}^{\zmax} dz\, P_{ab}(z),
\end{equation}
where $P_{ab}(z)$ are the Altarelli-Parisi splitting functions corresponding to the processes depicted in Fig.~\ref{fig:gluonVsQuarkParisi}. Unless explicitly required, the $\zmin$ and $\zmax$ arguments will be omitted. If we start with a quark (or gluon), the fraction of secondary Lund planes with a gluon as a leading parton is, respectively (assuming $\zmin<\zmax$)
\begin{equation}\label{eq:g-frac-collinear-groom}
f^\text{(groom)}(g|q) = \frac{I^\text{(groom)}_{gq}}{I^\text{(groom)}_{qq}+I^\text{(groom)}_{gq}}
\qquad\text{ and }\quad
f^\text{(groom)}(g|g) = \frac{I^\text{(groom)}_{gg}}{I^\text{(groom)}_{qg}+I^\text{(groom)}_{gg}}.
\end{equation}
All integrals in these expressions can be computed analytically. 

The case of the dijet selection is slightly more involved. Indeed, since we are imposing separate dimensionful cuts on the leading and subleading jets, the selected rates in each partonic channel depend on the underlying jet cross-section. Still working in the collinear limit, we thus introduce
\begin{align}
\label{eq:int-Pab-dijet}
I^\text{(dijet)}_{ab} 
& = \int dp_t \frac{d\sigma_b(p_t)}{dp_t} \int_0^1 dz\, P_{ab}(z)
\,\Theta((1-z)p_t>p_{t,\text{lead}}^\text{(min)})
\,\Theta(p_{t,\text{sublead}}^\text{(min)}<zp_t<p_{t,\text{sublead}}^\text{(max)}),\nonumber\\
& = \int dp_t \frac{d\sigma_b(p_t)}{dp_t} I^\text{(groom)}_{ab}\left(\frac{p_{t,\text{sublead}}^\text{(min)}}{p_t},\min\left(1-\frac{p_{t,\text{lead}}^\text{(min)}}{p_t},\frac{p_{t,\text{sublead}}^\text{(max)}}{p_t}\right)\right),
\end{align}
where, at leading order in $\alpha_s$, the flavour decomposition of the underlying Born-level process is well-defined.
The fraction of secondary Lund planes with a gluon as a leading parton is 
\begin{equation}\label{eq:g-frac-collinear-dijet}
f^\text{(dijet)}(g|q) = \frac{I^\text{(dijet)}_{gq}}{I^\text{(dijet)}_{qq}+I^\text{(dijet)}_{gq}}
\qquad\text{ and }\quad
f^\text{(dijet)}(g|g) = \frac{I^\text{(dijet)}_{gg}}{I^\text{(dijet)}_{qg}+I^\text{(dijet)}_{gg}}.
\end{equation}
respectively for a leading quark or leading gluon.
It is interesting to notice that while the gluon fractions $f^\text{(groom)}$ are only functions of $\zmin$ and $\zmax$, the corresponding fractions  $f^\text{(dijet)}$ also depend on the choice of PDFs and of factorisation scale through the inclusive jet cross-section. 
In practice, we have found that this dependence is however very small, much smaller than the uncertainty associated with the actual fraction of quarks and gluons which also depends on the choice of PDFs and factorisation scale. 
The integrals $I^\text{(dijet)}_{ab}$ can only be computed analytically for a simple approximation of the inclusive jet cross-section. Since the resulting equations are not particularly helpful, the results presented below use the explicit integration with the same PDF set as the one used with \texttt{NLOJet++}.

In addition to this DGLAP-based estimate, we have also ran a MC simulation of Born-level $pp\to jjj$ events using an extended version of \texttt{NLOJet++} that has access to flavour information. The centre-of-mass energy is set to $\sqrt{s}=13.6$ TeV. We choose the \texttt{PDF4LHC21\_mc} PDF set and account for uncertainties using the Hessian reduction strategy~\cite{PDF4LHCWorkingGroup:2022cjn}. The central values for the factorisation and renormalisation scales are set to the scalar sum of the transverse momentum of the final-state partons. Scale variations are those provided by \texttt{NLOJet++} and we take the envelope to define the scale uncertainty. The total perturbative uncertainty of our exact fixed-order result is obtained by adding in quadrature the scale and PDF uncertainties. 

\begin{figure}
    \centering
  \includegraphics[page=4,width=0.32\textwidth]{figures/born-fractions.pdf}
 \includegraphics[page=5,width=0.32\textwidth]{figures/born-fractions.pdf}
 \includegraphics[page=6,width=0.32\textwidth]{figures/born-fractions.pdf} \\
     \includegraphics[page=1,width=0.32\textwidth]{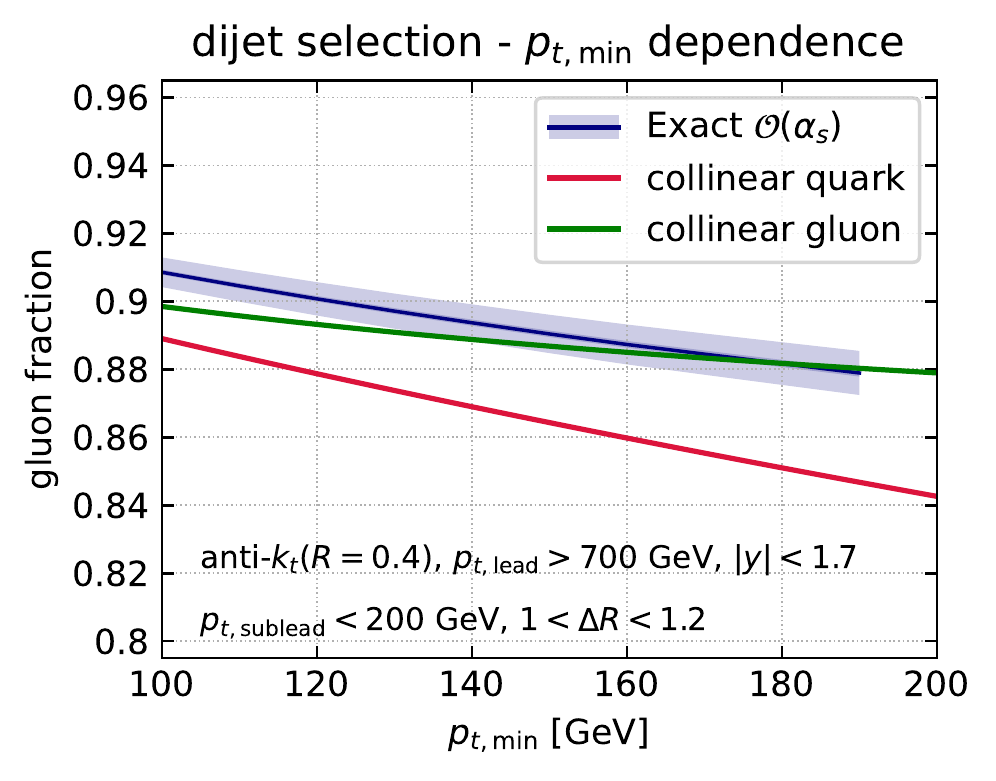}
   \includegraphics[page=2, width=0.32\textwidth]{figures/born-fractions.pdf}
   \includegraphics[page=3, width=0.32\textwidth]{figures/born-fractions.pdf} 
    \caption{Analytic estimates for the gluon fraction of the subleading subjet in the grooming setup (top row) and of the subleading jet with a dijet selection (bottom row). The different columns explore the dependence of the gluon fraction on the free parameters of each setup. On average, we achieve gluon fractions of around $90\%$.}
    \label{fig:gfrac-analytics}
\end{figure}

The obtained gluon fractions are shown in Fig.~\ref{fig:gfrac-analytics}. We first discuss the groomed selection. In this case, we choose large-radius jets ($R=1.2$) at high-$p_t$ ($p_t> 1$ TeV). We also impose a rapidity cut ($|y|<1.7$) to account for the acceptance of the CMS and ATLAS detectors. The free parameters in this setup are $\zmin$ and $\zmax$ together with the minimum opening angle $\Delta_{\rm min}$ between the prongs. Overall, the analytic estimate in the collinear limit captures the trends of the exact $\mathcal{O}(\alpha_s)$ result except for the $\Delta_{\rm min}$ dependence since Eq.~\eqref{eq:g-frac-collinear-groom} do not depend on any angular scale. The gluon fraction ranges from 87\% to 93\%. We observe that it decreases when increasing either $\zmin$ or $\zmax$ (keeping the other one fixed). The reduction is more pronounced for a quark-initiated jet. This decrease of the gluon fraction can be traced back to the fact that the ratio $P_{qx}(z)/P_{gx}(z)$ (with $x=q$ or $x=g$) increases for $0<z<1/2$, meaning that an increasingly large fraction of the secondary partons will be quarks when one departs from the soft limit where $P_{qx}(z)/P_{gx}(z)=0$.
In turn, the $\Delta_{\rm min}$-dependence is rather flat at this level of accuracy. The collinear estimates indicate that fixing $\zmin=0.1$ with $\zmax=0.2$ leads to almost identical gluon-fraction for quark or gluon initiated jets, around 91\%.   

In the dijet selection case, we analyse jets with standard anti-$k_t$ properties ($R=0.4$). The leading jet is again at high-$p_t$ ($p_t> 700$ GeV) and only jets with $|y|<1.7$ are accepted. The free parameters for this setup are the $p_t$-window for the subleading jet characterized by $\ptmin$ and $\ptmax$ together with the minimum angular separation between the leading and subleading jets, $\Delta R_{\rm min}$. Once again, the collinear estimate for the gluon fraction,  Eq.~\eqref{eq:g-frac-collinear-dijet}, correctly describes the trends of the exact fixed-order result. In this case, we find a milder dependence of the gluon fraction with respect to the different parameters. It stays around 90\% in the explored parameter space. Enlarging either $\ptmin$ or $\ptmax$ results into a reduction of the gluon fraction. For just one emission, $\ptmin$ or $\ptmax$ plays the same role as the momentum imbalance cuts, $\zmin$ with $\zmax$, which we discussed in the groomed setup. In other words, increasing $\ptmin$ is equivalent to increasing $\zmin$. Therefore, it is natural that the dijet and the groomed setups display a similar behaviour. Once again, we find that, at $\mathcal{O}(\alpha_s)$, the gluon fraction does not change when varying $\Delta_{\rm min}$.  

Because the dijet selection approach has the possibility of branching out to other applications beyond the secondary Lund jet plane, we will mainly focus on the latter for the remainder of this paper. The conclusions from the hadron-level simulation results discussed in the next section also hold for the SoftDrop and trimming setups.

\begin{figure}
    \centering
    \includegraphics[width = 0.49\textwidth]{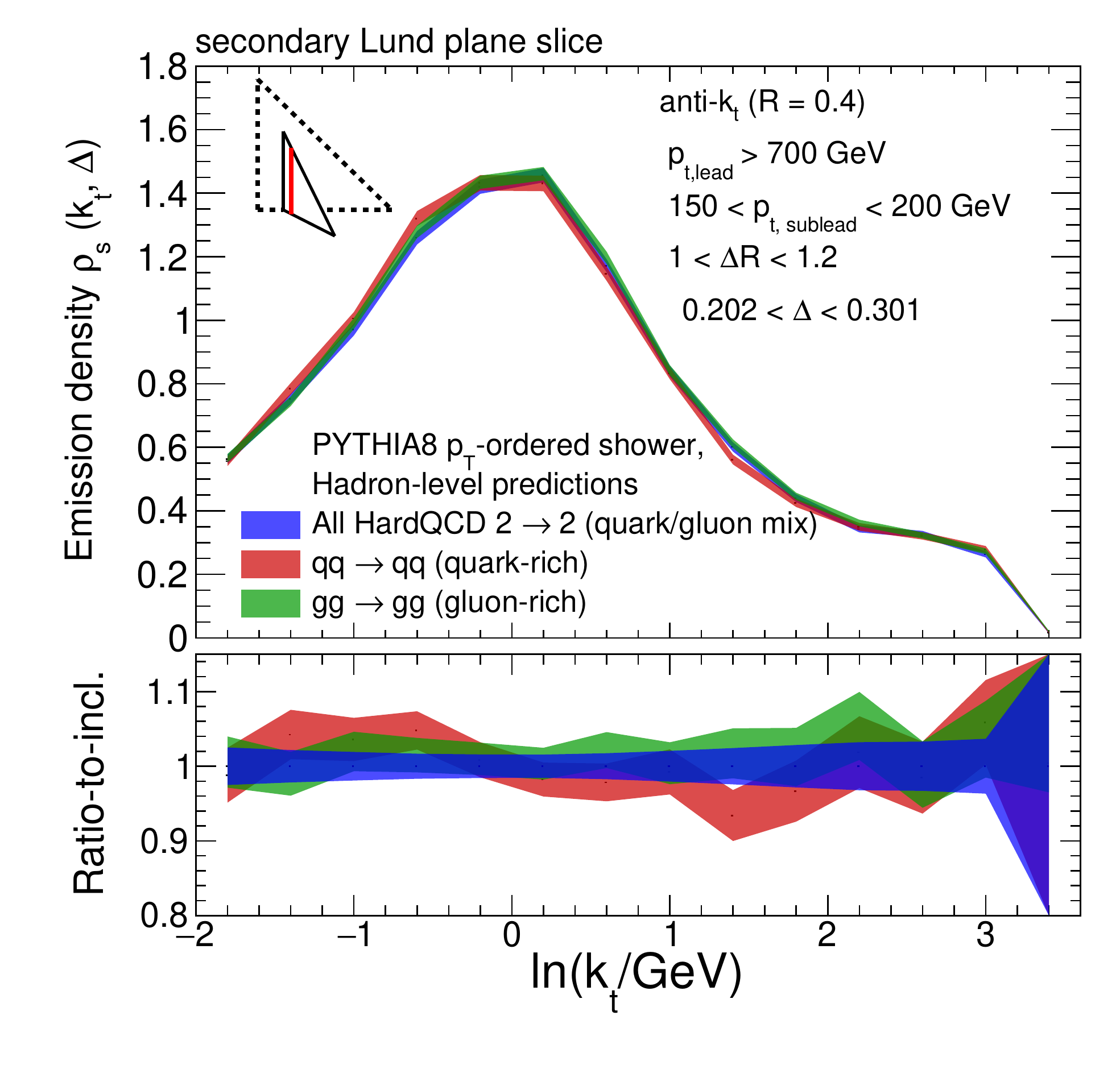}
    \includegraphics[width = 0.49\textwidth]{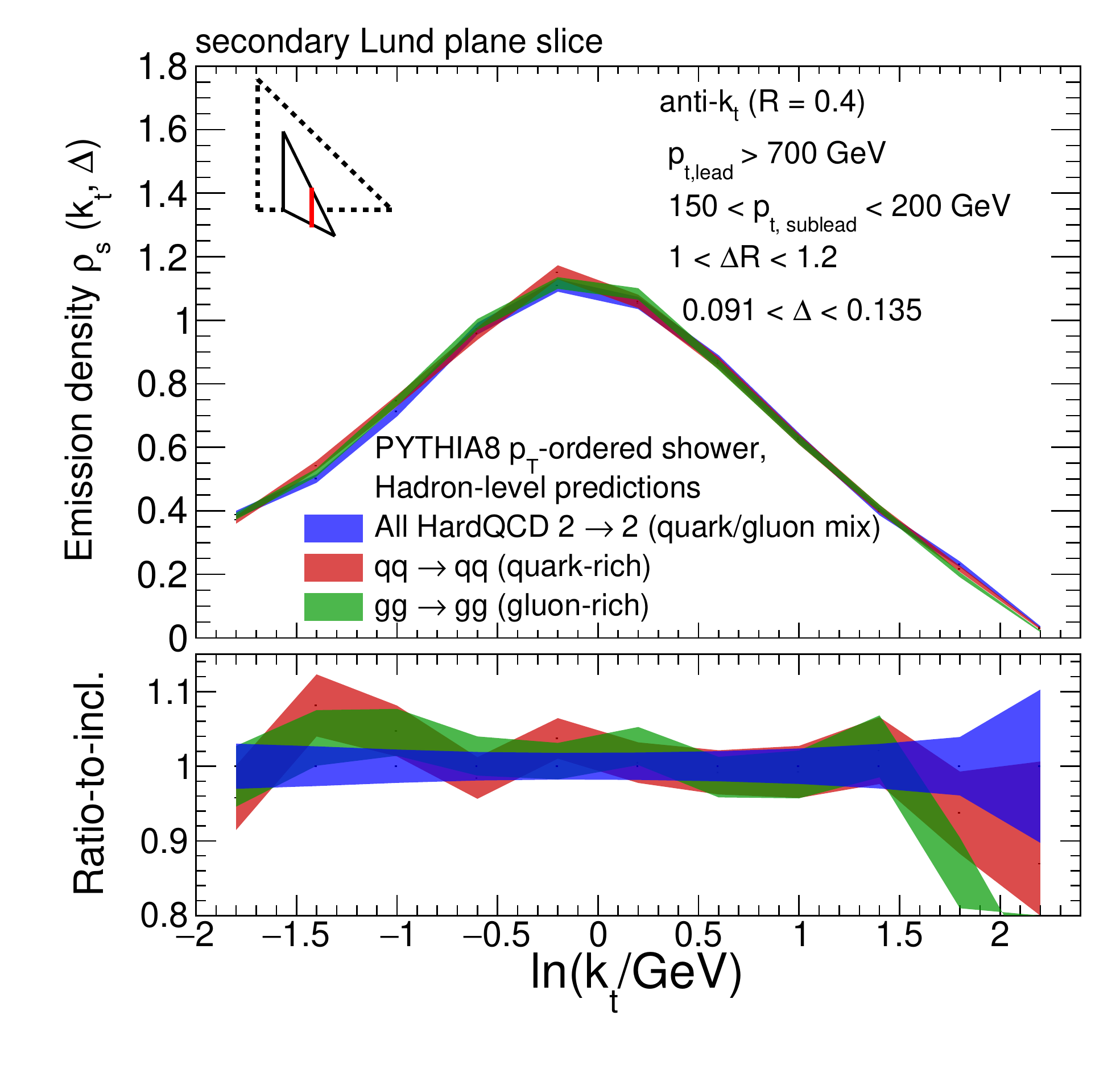}
    \caption{Vertical slices of the secondary Lund jet plane density at large (left) and small (right) angles. The three curves in each plot correspond to different choices for the hard $2\to 2$ scattering: all possible QCD channels (blue), only $qq\to qq$ (red) and $gg \to gg$ (green). The lower panel in each plot shows the ratio to the inclusive case. Results are shown for \py but similar conclusions can be drawn from \hw.}
    \label{fig:universality}
\end{figure}

\section{High-gluon purities at the LHC using a dijet selection}
\label{sec:MCs}

To match the expected amount of high-$p_t$ jet events collected in Run 2 with the CMS or ATLAS experiments, predictions are generated for 5 million events (similar amount of events are expected for Run 3). Only jet events produced purely by the strong interaction are considered, i.e., $2\to 2$ QCD scatterings. Photon bremsstrahlung off quarks is allowed in these simulation studies.\footnote{Occasionally, a large angle, hard photon emission from a quark will be selected as an emission to fill the secondary Lund jet plane, but these events are rejected in an experimental context with jet identification criteria and they represent a contribution at the per mille level in our MC-simulation studies.} The simulation is done with three different MC setups: \py with its default $p_t$-ordered shower~\cite{Bierlich:2022pfr}, and \hw with both the angular-ordered shower and its dipole variant~\cite{Bellm:2019zci}. In all cases, we present results at hadron-level with default settings (including the simulation of the underlying event, multiple parton interactions, colour reconnections and initial state radiation). Jets are clustered using the anti-$k_t$ algorithm and reclustered with the C/A algorithm using the FastJet package~\cite{Cacciari:2005hq,Cacciari:2011ma}. From these events, about $\mathcal{O}(10^5)$ gluon-like jets are obtained using the event selection requirements described above to select the primary emission, which corresponds to roughly $10^6$ Lund emissions filling the secondary Lund plane. In all figures presented in this section, the bands represent the expected statistical uncertainties at the LHC with these selections.

\begin{figure}
    \centering
    \includegraphics[width = 0.49\textwidth]{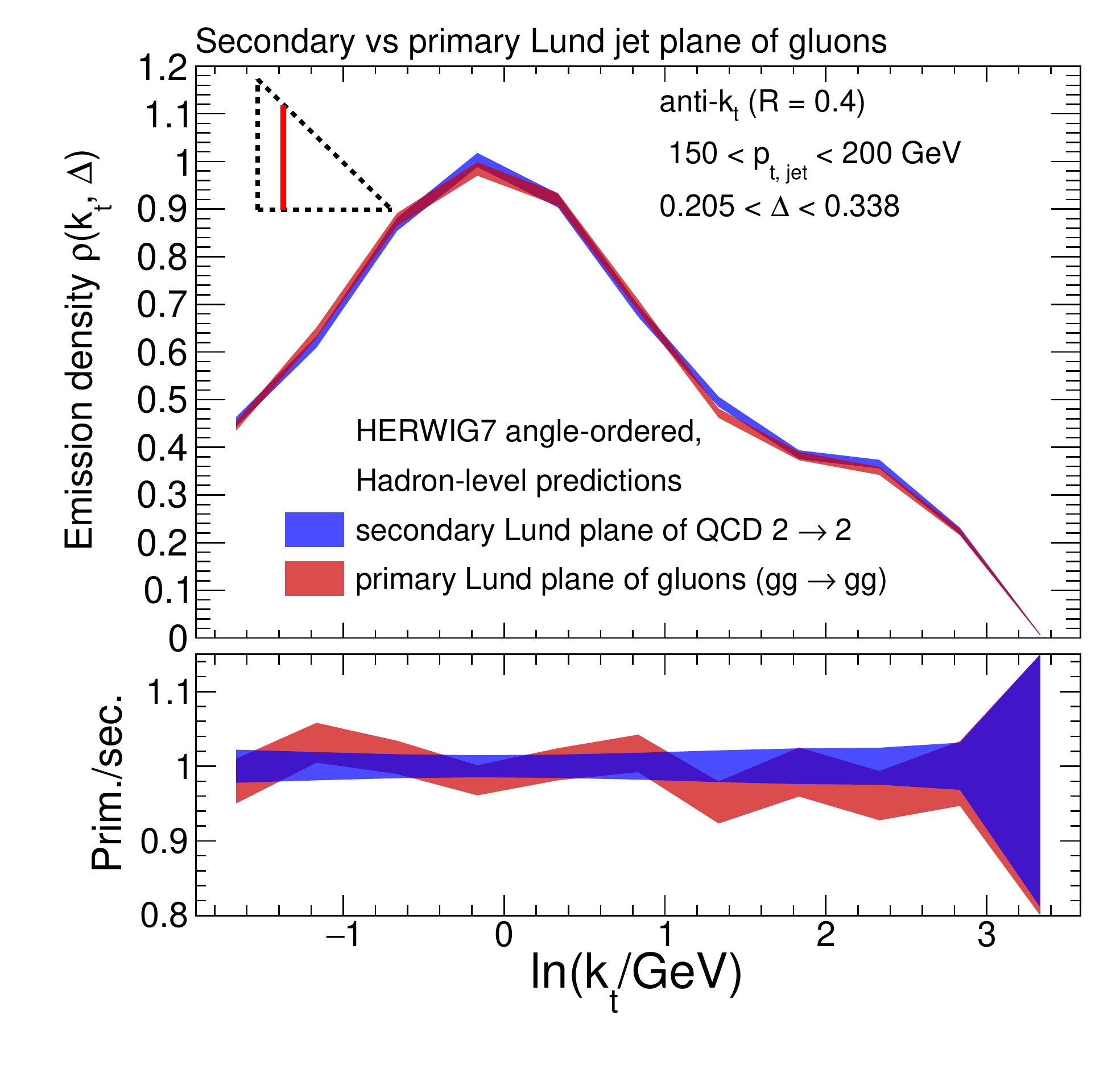}
    \includegraphics[width = 0.49\textwidth]{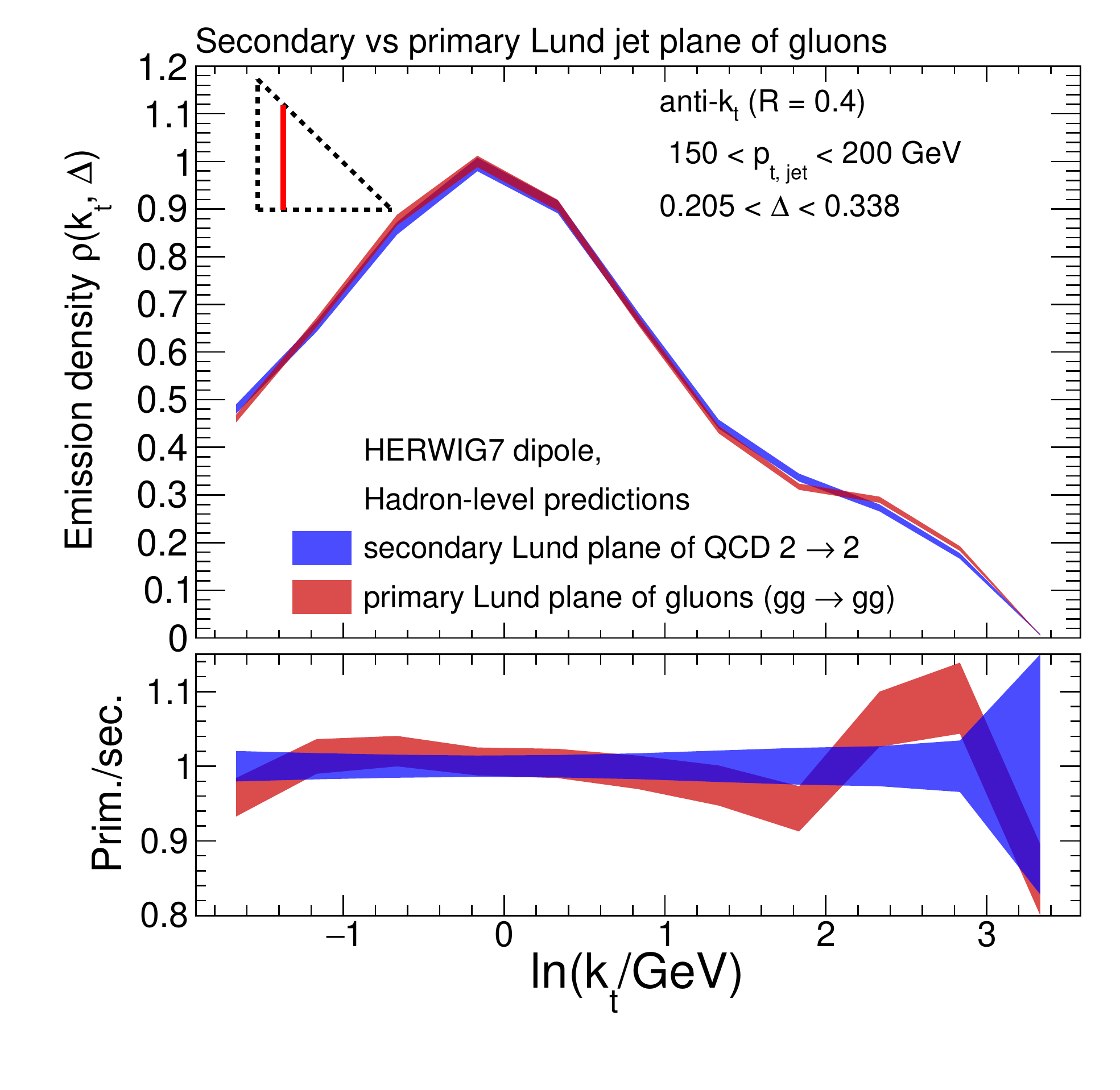}
    \caption{Vertical slice of a secondary Lund jet plane density derived from an inclusive jet sample (blue) compared with the primary Lund jet plane of jets produced in $gg\to gg$ scatterings (red). Both panels are generated with \hw either using the angle-ordered shower (left) or the dipole shower (right). The bottom panels show the ratio between the primary and secondary Lund plane densities.}
    \label{fig:secLP-vs-gg}
\end{figure}

First, we evaluate the dependence of $\rho_s$ on the hard-scattering process itself. That is, we compute $\rho_s$ either inclusively or fixing the $2\to 2$ scattering to be quark-rich ($qq\to qq$) or gluon-rich ($gg\to gg$). Despite being an academic exercise, it is important to demonstrate that the proposed methodology is independent of the flavour of the partons participating in the hard scattering. This also tests the sensitivity to the PDFs. The results are shown in Fig.~\ref{fig:universality} by choosing two angular slices of the secondary Lund plane density as obtained with \py. First we note that $\rho_s$ is peaked around $k_t=1$ GeV for this jet selection. Changes on the hard-scattering process translate into less than 10\% differences at the level of the secondary Lund plane density. This nice feature shows up both at large and small angles. Also, we find quantitative agreement between all MC generators discussed above.    

To assess if the secondary Lund jet plane yields the desired results (i.e., a sample of gluon-enriched radiation), it is useful to compare it with a reference distribution. For this purpose, we use the primary Lund jet plane of gluon-initiated jets generated from $gg\to gg$ scatterings. The reference primary Lund jet plane is built using the two highest $p_t$ jets in the event, with each of them required to have $150 < p_{t} < 200$ GeV and $|y| < 1.7$, i.e.\ the same $p_t$ and $y$ requirements that are used for the jets used for the secondary Lund jet plane analysis.\footnote{At the simulation level, the partonic hard scale, $\hat{p}_t$, has to be lower for the reference sample than the $\hat{p}_t$ used for the secondary Lund jet plane.}

We compare the secondary Lund plane density obtained from inclusive jets to the primary Lund plane density for gluon-initiated jets in Fig.~\ref{fig:secLP-vs-gg}. In this case we show results generated with the two parton showers of \hw. The curves agree within 10\% deviations throughout the whole $k_t$-range for both parton shower models. Thus, we find that $\rho_s$ is an excellent proxy for artificially created sample of pure gluon-initiated jets.

\begin{figure}
    \centering
    \includegraphics[width = 0.49\textwidth]{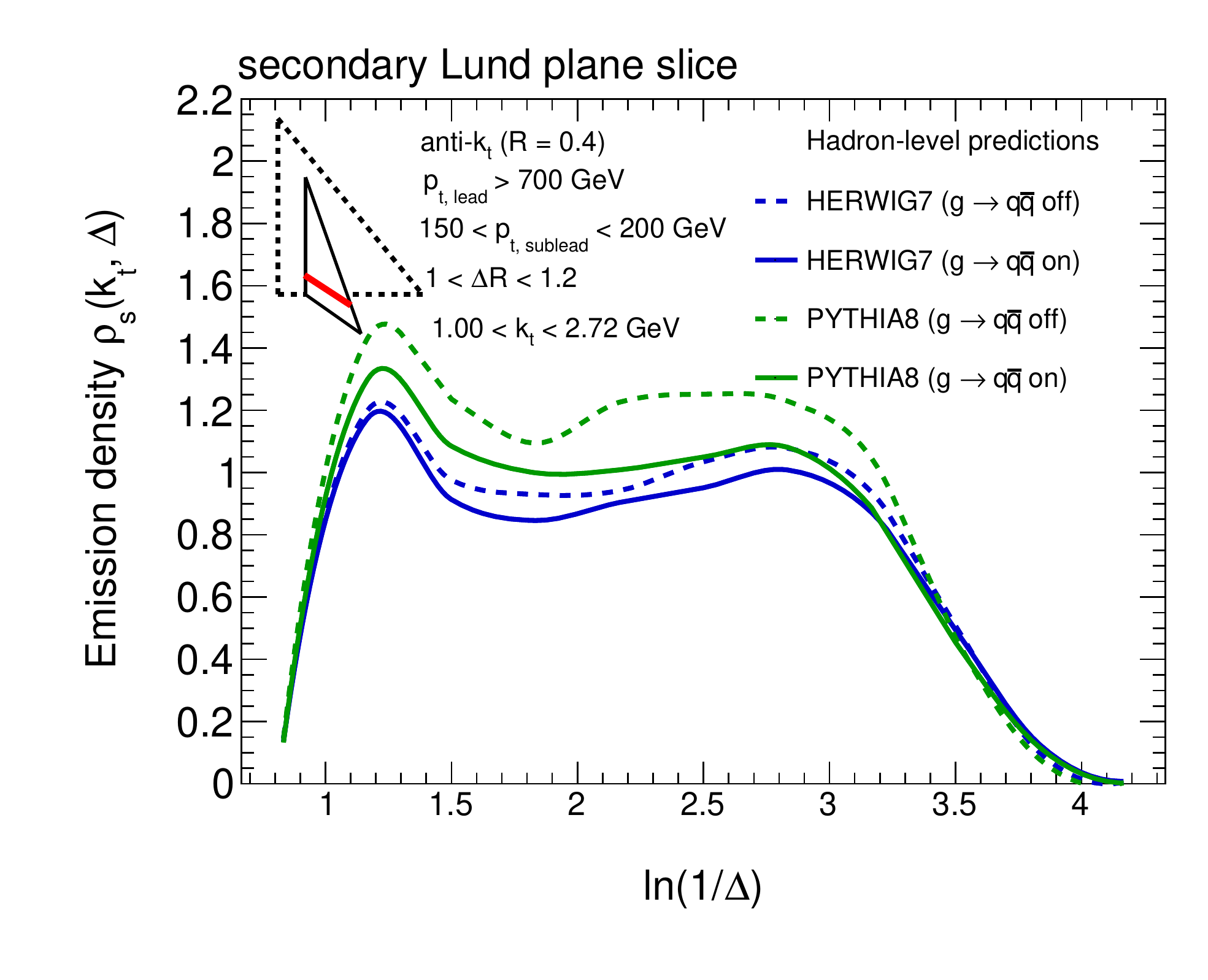}
        \includegraphics[width = 0.49\textwidth]{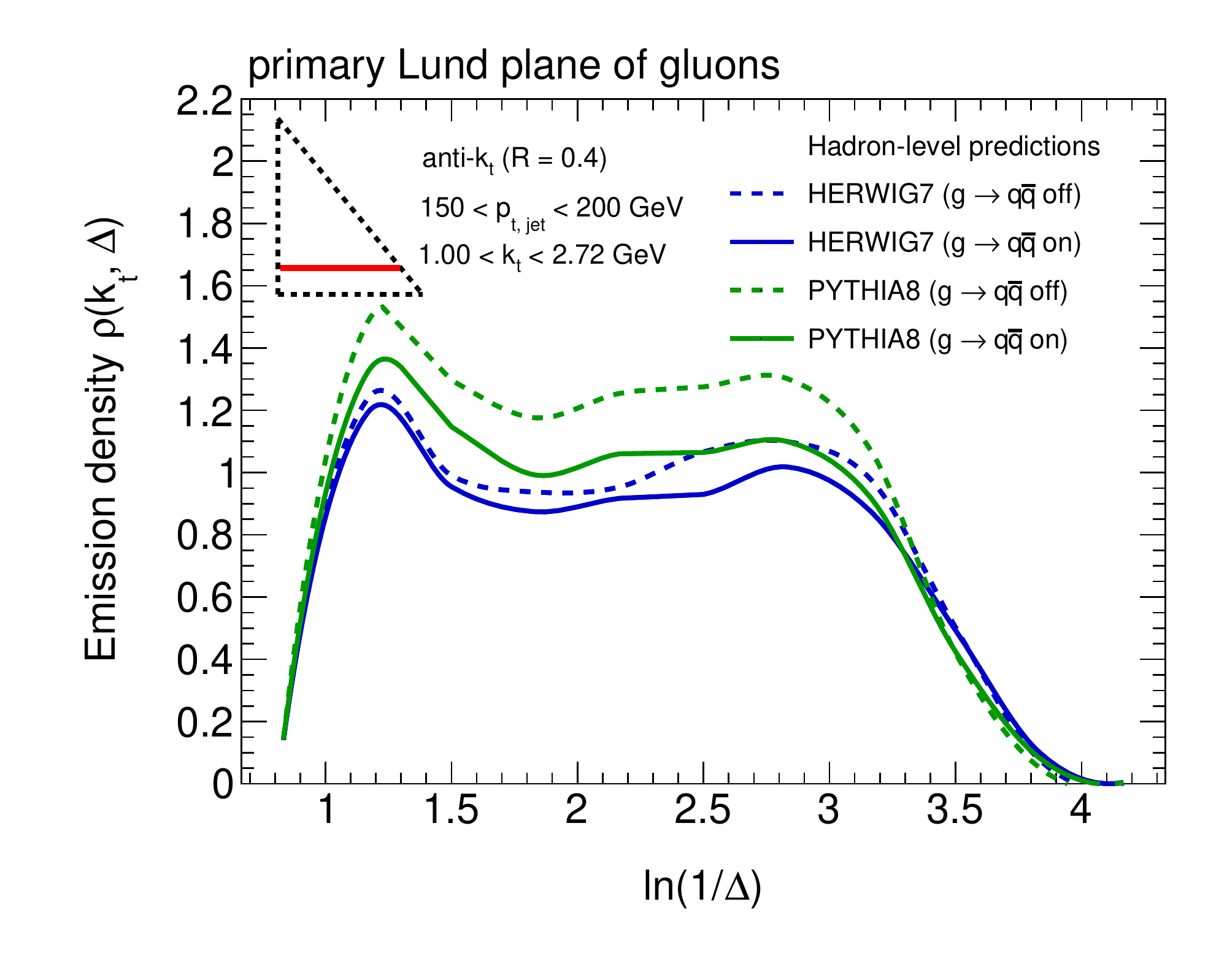}
    \caption{Impact of $g\to q\bar q$ splittings on a horizontal slice of the secondary Lund plane density (left) and of the primary Lund plane density from $gg\to gg$ scatterings (right). Results are shown for both \py and \hw.}
    \label{fig:g2qqBarOff}
\end{figure}

Next, we analyze the impact of gluon splittings to quark-antiquark pairs by switching them off in the simulation. Although not realistic from a physical point of view, this exercise is instructive to understand whether the proposed fiducial cuts manage to suppress this channel. We again use the primary Lund plane of gluon-initiated jets as our baseline to compare the performance of the secondary density. The results are displayed in Fig.~\ref{fig:g2qqBarOff} for a slice in transverse momentum. Removing $g\to q\bar q$ splittings leads to an increase of emissions in both \py and \hw (more pronounced for the former) for the primary Lund plane of gluons. By examining the secondary Lund plane, similar modifications are found at a quantitative level. In particular, the secondary Lund plane density remains almost identical to the primary density of gluon-initiated jets with our without $g\to q\bar q$ splittings. This further confirms that parton flavour changes from $g\to q\bar{q}$ are negligible with the proposed dijet selection and that the phenomenological implications of $g\to q\bar{q}$ splittings for gluon-initiated jets can also be constrained via the secondary Lund jet plane density.

\begin{figure}
    \centering
    \includegraphics[width = 0.49\textwidth]{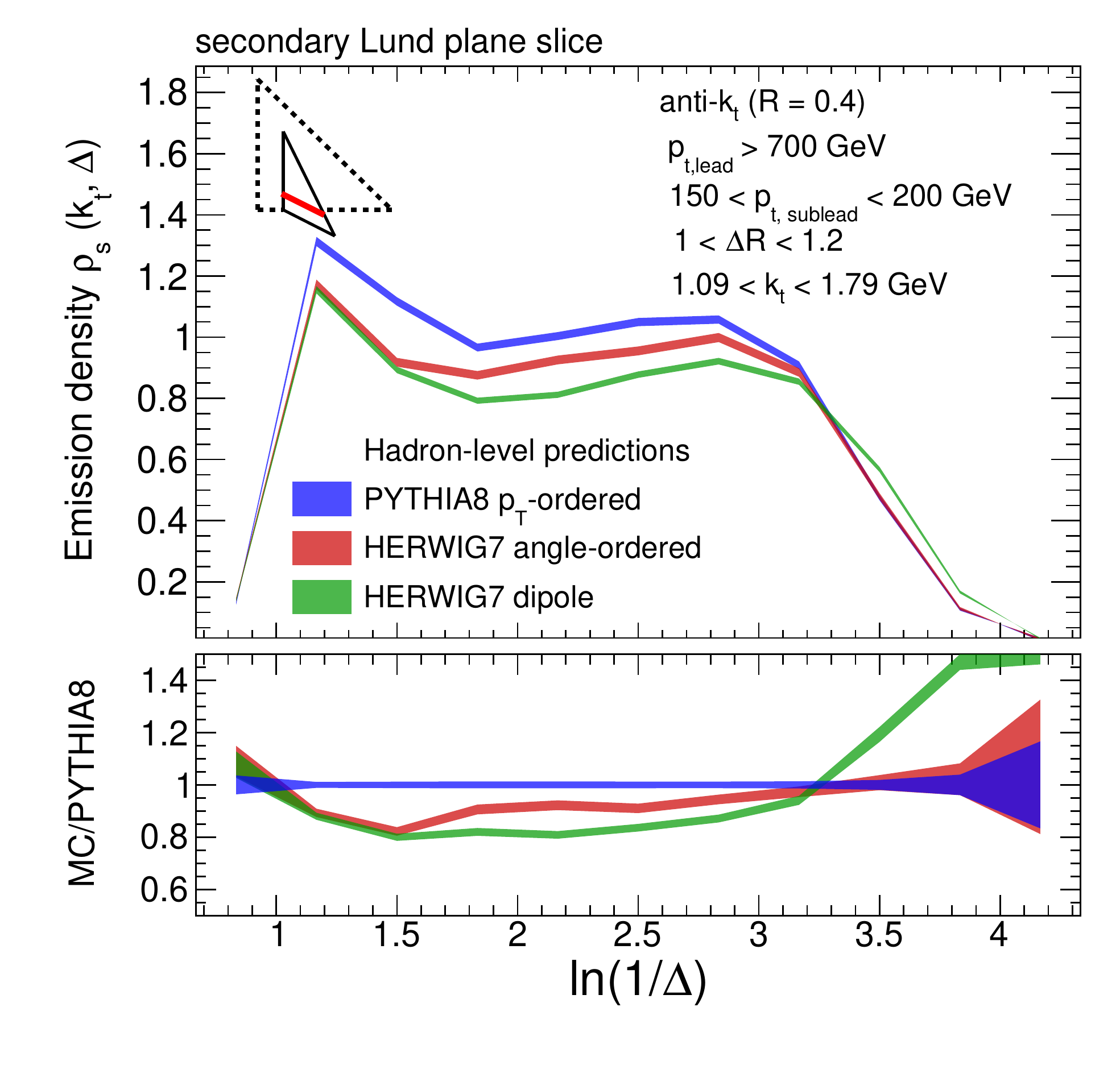}
        \includegraphics[width = 0.49\textwidth]{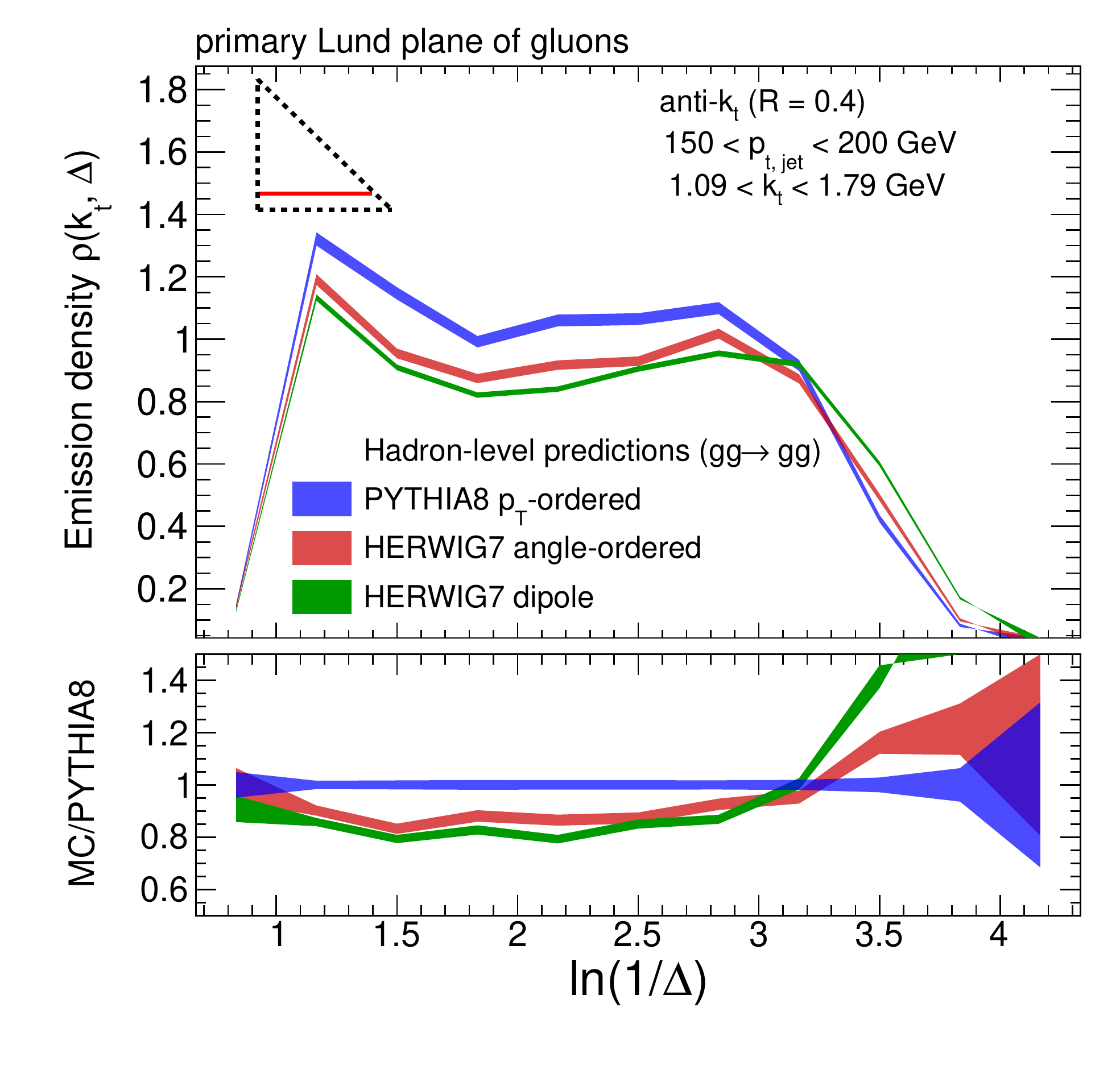}
    \caption{Comparison between different MC generators for a horizontal slice of the secondary Lund plane density (left) and of the primary Lund plane density for gluon-initiated jets (right). The bottom panel shows the ratio to the \py results.}
    \label{fig:mc-comparison}
\end{figure}

To end up this section, we explore the constraining power of the secondary Lund plane density in terms of MC generator models. We compare $\rho_s$ to the primary Lund plane density of gluon-initiated jets in a horizontal Lund plane slice in Fig.~\ref{fig:mc-comparison}. This comparison is meant to illustrate the extent to which one can discriminate parton showers and hadronisation models using only secondary Lund jet planes compared to an idealized scenario where one could unambiguously tag gluon-initiated jets. We find that the three MC setups start to differ in the deep collinear region, i.e., $\ln(1/\Delta)>3$. In this region of phase-space, the \hw dipole shower predicts a larger emission density for collinear emissions, almost a factor of 2 larger with respect to \hw with an angle-ordered shower or \py. The same trend is observed using the reference primary Lund plane for gluon jets. This shows that the secondary Lund plane densities matches the constraining power of a pure gluon-initiated jet sample. In other words, the spread among the different MC-generated predictions for secondary Lund jet planes is generally the same as the one observed for gluon-jet primary Lund jet plane densities.

\section{Summary and Outlook}
\label{sec:end}
In this paper, a proposal for extending the exploration of the Lund jet tree in high energy proton-proton collisions at the LHC has been presented. Specifically, the possibility of using the average density of emissions off a primary emission, the secondary Lund jet plane, to constrain gluon-initiated radiation is investigated. Our strategy heavily relies on a fundamental property of QCD: the soft divergence of QCD splitting functions ($q\to qg$ and $g\to gg$). That is, collinear gluon emissions will typically carry a small fraction of its parent's energy. As such, if one focuses on splittings with asymmetric momentum balance and selects the less energetic branch, the probability of it being a gluon is enhanced. 

We discuss three different ways of constructing the secondary Lund jet plane density of emissions, either with well-established jet substructure techniques (SoftDrop grooming, trimming) or with a dijet (multijet) event selection. Given the simplicity of the latter, we focus on it for most of the paper. Analytically, we demonstrate by means of fixed-order calculations that an inclusive dijet selection with a pair of high-$p_t$ jets sharing asymmetric momentum balance and collinear to each other can yield subleading jet gluon fractions that are, on average, around 90\% for realistic LHC fiducial cuts. This result is independent of the flavour of the jet-initiating parton to a large extent. Higher gluon purities can be reached by selecting softer subleading (sub)jets, at the expense of phase space reduction.

To further confirm the potential of such an observable, we perform hadron-level simulation studies with \hw and \py. We select anti-$k_t$ jets with $R=0.4$ and pair them up imposing that the leading jet has $p_{t,{\rm lead}}>700$ GeV and the subleading $150<p_{t,{\rm sublead}}<200$ GeV. On top of the $p_t$-selection, we also constrain their angular separation to be $1<\Delta R < 1.2$. For Run-2 luminosity, this corresponds to $\mathcal{O}(10^5)$ jets. The secondary Lund plane density is then constructed using the subleading jet. 

We find that the resulting secondary Lund plane densities are resilient, within $10\%$, to the flavour of the partons participating in the hard-scattering process (including their PDFs). Remarkably, the secondary Lund plane density agrees with the primary Lund plane density from gluon-initiated jets again within 10\%. We also study the sensitivity to flavour changes due to gluon to quark-antiquark splittings and find it to be negligible for our jet selection. Finally, the model discrimination power for secondary Lund jet plane densities is similar to the one obtained in an ideal primary Lund jet plane density from gluon-initiated jets. This supports the idea of using the secondary Lund plane density as a tool to constrain the modeling of gluon-initiated radiation.

Having access to a gluon-enriched sample using simple hadron-level phase space cuts opens up a series of interesting opportunities. At the phenomenological level, a measurement of the secondary Lund plane density can serve as a much needed input for constraining gluon-initiated radiation in MC event generators. This could potentially reduce the scale factor uncertainties for quark-gluon likelihood discriminators. Measurements of the hadron species composition and their energy distribution within jets in such a gluon-rich sample could serve as a guide for the reduction of uncertainties associated to the jet flavor response uncertainty for the jet energy scale. Regarding precision physics, suppressing the sensitivity to the quark/gluon jet fraction represents a major step forward towards the goal of extracting $\alpha_s$ from jet substructure observables. To this end, there are better candidates than the secondary Lund jet plane density from the theoretical perspective. For example, observables such as the SoftDrop groomed jet mass, or the average Lund multiplicities have a simpler resummation structure. The drawback of computing observables on the secondary Lund plane is that it pushes the accuracy of the fixed-order result to one order higher. For instance, providing a theoretical prediction that could result into a competitive $\alpha_s$ extraction would require the calculation of $pp\to jjj$ at least at NNLO~\cite{Czakon:2021mjy,ATLAS:2024png}, for which no public code currently exists. The complexity on the resummation remains the same as for observables on the primary branch. Nevertheless, we hope that the resilience of observables based on the secondary Lund plane to quark/gluon fractions and PDF uncertainties motivates the community to perform theoretical computations in this direction. In fact, the resilience of our method to the jet-initiating flavour can be readily tested experimentally using existing quark/gluon discriminators.

Another interesting spin-off of this work would be to apply this technique to heavy-ion jets. There, controlling the quark/gluon fraction represents a fantastic opportunity to understand the colour-charge dependence of energy loss. Several proposals to disentangle quark and gluon jets have been put forward in the literature~\cite{Chien:2018dfn,Li:2019dre,Brewer:2020och,Brewer:2021hmh,Ying:2022jvy,CMS:2020plq,Zhang:2023oid}. Despite the simplicity of our approach, its applicability to a heavy-ion context faces some challenges. One of them involves mitigating the underlying event activity that has so far hampered Lund plane measurements in heavy-ion collisions. In addition, a large sample of high-$p_t$ jets is required so as to guarantee sufficient phase-space to fill up the secondary Lund plane density. Heavy-flavor tagged jets can potentially be used to overcome some of these challenges, for instance extending the dijet selection to a trijet selection with two $b$-tagged jets and an anti-$b$ tagged third jet.

In summary, we have presented a simple strategy to measure a high-purity sample of gluon-initiated jets that (i) can be readily applied at the LHC, (ii) has a well-defined connection to pQCD calculations, and (iii) can be easily implemented in a Rivet routine, which greatly improves the ability of measurements based on this method to constrain Monte Carlo generators.

\acknowledgments
GS and ASO thank the CERN theoretical department for the hospitality at different stages of this work. CB thanks Laboratoire Leprince-Ringuet and Sapienza Universit\`{a} di Roma for the support provided in the development of this work. CB thanks Jacob March for his contributions at the early stages of this work, and Leticia Cunqueiro and Matthew Nguyen for helpful discussions. This work has been supported by the European Research Council (ERC) under the European Union’s Horizon 2020 research and innovation programme (grant agreement No. 788223, PanScales) and by the ERC consolidator programme (grant agreement No. 1010022070, QCDHighdensityCMS). ASO is supported by the Ramón y Cajal program under grant RYC2022-037846-I. CB is supported by the U.S. Department of Energy, Office of Science, Office of Nuclear Physics under grant contract number DE-SC0011088.
\bibliographystyle{elsarticle-num}
\bibliography{biblio.bib}
\end{document}